# Successive Phase Transitions under High Pressure in FeTe$_{0.92}$

Hironari OKADA[1,4*], Hiroyuki TAKAHASHI[1], Yoshikazu MIZUGUCHI[2,3,4], Yoshihiko TAKANO[2,3,4], and Hiroki TAKAHASHI[1,4]

[1]*Department of Physics, College of Humanities and Sciences, Nihon University, Sakurajosui, Setagaya-ku, Tokyo 156-8550, Japan*
[2]*National Institute for Materials Science, 1-2-1 Sengen, Tsukuba 305-0047, Japan*
[3]*University of Tsukuba, 1-1-1 Tennodai, Tsukuba 305-8577, Japan*
[4]*JST, Transformative Research Project on Iron Pnictides (TRIP), 5 Sanbancho, Chiyoda, Tokyo 102-0075, Japan*

We performed magnetization and electrical resistivity measurements under high pressures of up to 19 GPa for FeTe$_{0.92}$. The compound shows an anomaly in magnetization and resistivity at atmospheric pressure due to a structural distortion accompanied by a magnetic transition. We also observed magnetic and resistive anomalies under high pressure, suggesting that two pressure-induced phases exist at a low temperature. Unlike in FeAs-based compounds, no superconductivity was detected under high pressures of up to 19 GPa, although the anomaly at atmospheric pressure was suppressed by applying pressure.

KEYWORDS: iron-based superconductor, FeTe, high pressure, electrical resistivity, superconductivity

Soon after the discovery of FeAs-based superconductors with a superconducting transition temperature $T_c$ of up to 55 K,[1-5] superconductivity with $T_c$ = 8 K in tetragonal FeSe$_x$ was discovered.[6] Similarly to FeAs-based superconductors, the tetragonal FeSe$_x$ has edge-sharing FeSe$_4$ layers and its crystal structure is composed of a stack of Fe$_2$Se$_2$ layers along the $c$-axis. The $T_c$ of FeSe$_x$ is increased by the substitution of Se with S and Te.[7-9] In contrast to FeSe$_x$, isostructural Fe$_{1+x}$Te shows no superconductivity but exhibits a first-order structural phase transition accompanied by an antiferromagnetic (AFM) transition.[10,11] Depending on the amount of excess Fe, $x$, Fe$_{1+x}$Te has different crystal and magnetic structures at a low temperature despite the same tetragonal structure at room temperature. The compound with a nearly stoichiometric composition exhibits a monoclinic distortion and a commensurate AFM ordering at low temperature, whereas orthorhombic and incommensurate AFM structures are realized for a composition with a large amount of excess Fe. The incommensurate AFM ordering locks into the commensurate AFM ordering with decreasing amount of excess Fe. Although the structural distortion and magnetic structures at a low temperature are different from those observed for FeAs-based compounds,[12-16] the structural and magnetic phase transitions are suppressed by the substitution of Te with S and Se, similarly to those in FeAs-based compounds, and then a superconducting transition with $T_c$ = 10-14 K occurs.[7,8,17-20]

Superconductivity in FeAs-based compounds is very sensitive to external pressure. Not only the substitution effect, but also external pressure affects the occurrence of superconductivity in parent compounds of FeAs-based superconductors, which results from the suppression of the structural phase transition from tetragonal to orthorhombic and of AFM ordering by the application of pressure.[21-25] The superconductivity of FeSe$_x$ is also sensitive to external pressure and shows a marked increase in $T_c$, which reaches ~37 K at 7-9 GPa.[26-29] This pressure effect of FeSe$_x$ is larger than that of LaFeAsO$_{1-x}$F$_x$.[30] A previous study on FeTe$_{0.92}$ at high pressures revealed that the resistive anomaly due to structural and magnetic phase transitions tends to be suppressed by applying pressure, although the applied pressure is limited to less than 1.6 GPa.[31] In addition, a theoretical investigation on stoichiometric FeS, FeSe, and FeTe predicted that the Fermi surface of these compounds is very similar to that of FeAs-based compounds, and that both FeSe and FeTe show a spin-density-wave ground state.[32] The instability of the spin-density-wave in FeTe is greater than that in FeSe. Therefore, it is expected that superconductivity with higher $T_c$ than that of FeSe$_x$ is realized under higher pressure in FeTe. In this study, we performed magnetization and electrical resistivity measurements under high pressures of up to 19 GPa for FeTe$_{0.92}$. We detected no signature of pressure-induced superconductivity but found pressure-induced successive phase transitions in FeTe$_{0.92}$.

A single crystal with nominal composition FeTe$_{0.92}$ was grown using a melting method. A polycrystalline FeTe$_{0.92}$ sample,[30] which is the starting material for single crystals, was sealed in an evacuated quartz tube with an alumina crucible. The sample was heated at 1123 K and slowly cooled to 1053 K at a rate of 1 K/h. Magnetization measurements under high pressures of up to 1.40 GPa were performed using a superconducting quantum interference device (SQUID) magnetometer (Quantum Design) and a piston-cylinder-type cell made of CuBe with a liquid pressure-transmitting medium (Daphne 7373). The applied pressure in the magnetization measurements was estimated from $T_c$ using a tin manometer. Electrical resistivity along in-plane direction under high pressure is measured by a standard dc four-probe technique. Pressures of up to 2.5 GPa were applied using a piston-cylinder-type (CuBe/NiCrAl) cell. A liquid pressure-transmitting medium (Daphne 7474) was used to maintain hydrostatic condition.[33] The applied pressure was estimated from $T_c$ using a lead manometer. A diamond anvil cell (DAC) was used for electrical resistivity measurements under high pressures of up to 19 GPa. The sample chamber equipped with a stainless-steel gasket was filled with powdered NaCl as a pressure-transmitting medium. Fine ruby powder scattered in the sample chamber was used to determine the applied pressure by a standard ruby fluorescence method.

Figure 1 shows the temperature dependence of the magnetization $M(T)$ at various pressures of up to 1.40 GPa. The magnetization at 0 GPa shows a jump at $T_s = 70$ K, which is due to the structural phase transition accompanied by the magnetic transition. $T_s$ decreases with increasing pressure, but the jump was observed up to 1.40 GPa, indicating that the structural phase transition occurs under high pressures of up to 1.40 GPa. On the other hand, the magnetization below $T_s$ above 1.14 GPa gradually decreases with decreasing temperature and changes to a weak temperature dependence below $T_0$, as indicated by arrows in Fig. 1. The weak temperature dependence at a lower temperature probably originates from the magnetization arising from the pressure cell. The gradual decrease in $T_0 \leq T \leq T_s$ above 1.14 GPa is not observed below 0.98 GPa, while the data below $T_0$ are similar to those below $T_s$ at 0 GPa. This result suggests that a pressure-induced magnetic phase appears in the intermediate-temperature region ($T_0 \leq T \leq T_s$) above 1.14 GPa, and that the magnetic transition at 0 GPa is suppressed by external pressure.

Figures 2(a) and 2(b) show the temperature dependences of the electrical resistivity $\rho(T)$ and $d\rho/dT$ below 160 K at various pressures of up to 2.5 GPa, respectively. The data at 0, 0.4, 1.1, 1.3, and 1.6 GPa in Fig. 2(a) are shifted upward by 3.5, 3.2, 2.5, 1.5, and 0.5 μΩm, respectively. $\rho(T)$ at 0 GPa slightly increases with decreasing temperature and shows a sudden decrease at $T_s$ = 69 K, which is consistent with the result of the magnetization measurements. Below $T_s$, resistivity rapidly decreases with decreasing temperature. $\rho(T)$ at 0 GPa is similar to that of $Fe_{1-x}Te$ with a nearly stoichiometric composition rather than to that of a compound with a large amount of excess Fe having semiconductive behavior in $\rho(T)$.[7,10,19] Therefore, it is considered that the sudden decrease at $T_s$ is due to the monoclinic distortion accompanied by the commensurate AFM ordering. $T_s$ shifts to a lower temperature at 0.4 GPa. At 1.1 GPa, $\rho(T)$ shows two anomalies at 55.4 and 62.2 K. These anomalies are also seen in $d\rho/dT$, as shown in Fig. 2(b). The two anomalies correspond to $T_0$ and $T_s$ observed in the magnetization measurements, respectively. $T_0$ and $T_s$ are also observed at 1.3 GPa, but no anomaly at $T_0$ is detected at 1.6 GPa. By further applying pressure, in addition to the anomaly at $T_s$, the $\rho(T)$ at 1.8 GPa exhibits an anomaly showing a large drop at $T^*$ = 45 K. $T^*$ shifts upward with increasing pressure, and no $T_s$ is detected above 2.3 GPa.

$\rho(T)$ below $T_0$ is similar to that at 0 GPa, whereas $\rho(T)$ below $T_s$ above 1.1 GPa shows a gradual decrease with decreasing temperature, leading to an increase in residual resistivity. The pressure dependences of the resistivities at 4.2 and 200 K are shown in Fig. 3. The resistivity at 200 K monotonically decreases with increasing pressure. On the other hand, the resistivity at 4.2 K decreases in the lower-pressure region but increases above 1.0 GPa. By further applying pressure, the resistivity at 4.2 K rapidly decreases above ~1.5 GPa. These results indicate that two high-pressure phases exist at a low temperature under high pressure. It is thought that these high-pressure phases are probably related to pressure-induced structural and magnetic phase transitions, and that the band structure and/or the scattering mechanism changes due to the phase transitions, resulting in different $\rho(T)$ curves below $T_s$, $T_0$, and $T^*$.

The pressure dependences of the characteristic temperatures $T_s$, $T_0$, and $T^*$ are shown in Fig. 4. $T_s$, which indicates the distortion from a tetragonal structure to a monoclinic structure, decreases with increasing pressure. This result is consistent with results of a polycrystalline sample in a previous report.[31] However, in our measurement using single crystals, we observed that $T_0$ appears at ~1.0 GPa and vanishes at ~1.5 GPa. The temperature dependences of the magnetization and resistivity below $T_0$ are similar to those below $T_s$ at 0 GPa, indicating that the crystal and magnetic structures below $T_0$ are the same as those below $T_s$ at 0 GPa, and that the commensurate AFM ordering is suppressed at ~1.5 GPa. In addition, the first high-pressure phase (HP I) exists between 1.0 and 2.3 GPa. It is considered that this high-pressure phase has a monoclinic structure but that the magnetic state is different from that at 0 GPa. In a higher pressure region, the second high-pressure phase (HP II) is induced above ~1.8 GPa below $T^*$. $T_0$ and $T^*$ are sensitive to external pressure, whereas $T_s$ has a weak pressure dependence. In particular, $T^*$ rises steeply with pressure at a rate of ~37 K/GPa, indicating that HP II below $T^*$ is rapidly stable to higher temperature under high pressure. Note that at 1.6 GPa the phase transition at $T_s$ occurs but no transition at $T_0$ or $T^*$ is detected.

Figure 5 shows the temperature dependence of electrical resistivity under high pressures of up to 19 GPa obtained using the DAC. A rapid decrease in electrical resistivity was observed in this measurement. This anomaly is probably due to the phase transition at $T^*$, although the anomaly observed in this measurement is broader than that in Fig. 1 owing to the nonhydrostatic compressive stress arising from the use of a solid pressure-transmitting medium. The rapid decrease in $\rho(T)$ shifts to a markedly higher temperature with increasing pressure and exceeds 300 K at 10 GPa. This result is consistent with the results obtained using the piston-cylinder-type cell. By further applying pressure above 10 GPa, $\rho(T)$ shows a negative temperature coefficient at a lower temperature. The origin of this behavior is unclear at present, but it may suggest that a further pressure-induced transition occurs above 10 GPa. In our measurements, no sign of pressure-induced superconductivity was detected even at high pressures of up to 19 GPa. Here, we used a stainless-steel gasket in this measurement, but its magnetism did not affect the observation of superconductivity.

From recent reports on FeAs-based compounds,[21-25] the parent compounds have an antiferromagnetically ordered orthorhombic phase at a low temperature, and superconductivity is induced by the suppression of both the orthorhombic structure and the AFM state through the substitution and/or pressure effect. Although the low-temperature phase at 0 GPa in FeTe$_{0.92}$, which has the monoclinic structure and commensurate AFM ordering, was suppressed by external pressure, we observed no signature of superconductivity under high pressures of up to 19 GPa. Instead, we found that pressure-induced successive phase transitions occur. HP I exists in the vicinity of 1.6 GPa and HP II is induced above ~1.8 GPa. The magnetic structures in the low-temperature phase of Fe$_{1-x}$Te are reasonably sensitive to the deviation in stoichiometric composition.[10,11] The commensurate AFM structure changes to an incommensurate AFM structure with increasing $x$. It is expected that this magnetic transition is induced by external pressure, which may be related to the appearance of HP I. HP II is rapidly stable to higher temperatures with increasing temperature and exceeds 300 K at 10 GPa. Recent X-ray diffraction experiments on FeSe under high pressure have revealed that the crystal structure changes from tetragonal to hexagonal above 8 GPa at room temperature.[27-29] The $T_c$ of FeSe shows a strong pressure dependence; it increases to ~37 K at 7-9 GPa, but decreases at higher pressure.[27-29] It seems that the hexagonal phase of FeSe is stable under high pressure and suppresses the superconductivity. Therefore, HP II below $T^*$ may be attributed to a structural phase transition resulting in the suppression of superconductivity.

A pressure-induced phase transition has also been reported in $CaFe_2As_2$, which is an FeAs-based compound having a $ThCr_2Si_2$-type tetragonal structure. $CaFe_2As_2$ exhibits a structural phase transition from tetragonal to orthorhombic at 0 GPa.[15] The compound also has a nonmagnetic collapsed tetragonal phase above ~0.35 GPa; this phase shows a marked reductions of 9.5% in the $c$-lattice parameter and 11% in the $c/a$ ratio in comparison with the orthorhombic phase.[34,35] The collapsed tetragonal phase is caused by the enhancement of the As-As bonds between neighboring FeAs layers under high pressure.[36] The crystal structure of $FeTe_{0.92}$ has a simple structure composed of only a stack of FeTe layers along the $c$-axis. Moreover, recent X-ray diffraction experiments on FeSe under high pressure have revealed that the compound shows the smallest bulk modulus of ~30 GPa and the largest compressibility along the $c$-axis among FeAs-based compounds.[27-29,37] Therefore, it is expected that $FeTe_{0.92}$ will also have a small bulk modulus and large anisotropic compression. Although details of the structural and magnetic properties under high pressure are still unclear at present, the Te-Te hybridization between neighboring FeTe layers is probably enhanced by applying pressure, resulting in the unique $P$-$T$ phase diagram shown in Fig. 4.

In summary, we performed the magnetization and the electrical resistivity measurements under high pressures of up to 19 GPa for $FeTe_{0.92}$. The low-temperature phase at 0 GPa with a monoclinic distortion and a commensurate AFM ordering was suppressed at ~1.5 GPa. However, we detected no pressure-induced superconductivity up to 19 GPa. Instead, we found that the anomalies at $T_0$ and $T^*$ appear at 1.1 and 1.8 GPa, respectively, suggesting the existence of two high-pressure phases. The first high-pressure phase exists in a narrow pressure region in the vicinity of 1.6 GPa. On the other hand, the second high-pressure phase below $T^*$ is rapidly stable to higher temperatures with increasing pressure and exceeds 300 K at 10 GPa. These successive phase transitions under high pressure are probably due to the enhancement of the Te-Te hybridization between neighboring FeTe layers.

**Acknowledgments**

This work was supported by a Grant-in-Aid for Young Scientists and JST-TRIP.

**References**

1) Y. Kamihara, T. Watanabe, M. Hirano, and H. Hosono: J. Am. Chem. Soc. **130** (2008) 3296.
2) X. H. Chen, T. Wu, G. Wu, R. H. Liu, H. Chen, and D. F. Fang: Nature **453** (2008) 761.
3) C. F. Chen, Z. Li, D. Wu, G. Li, W. Z. Hu, J. Dong, P. Zheng, J. L. Luo, and N. L. Wang: Phys. Rev. Lett. **100** (2008) 247002.
4) Z. A. Ren, G. C. Che, X. L. Dong, J. Yang, W. Lu, W. Yi, X. L. Shen, Z. C. Li, L. L. Sun, F. Zhou, and Z. X. Zhao: Europhys. Lett. **83** (2008) 17002.
5) M. Rotter, M. Tegel, and D. Johrendt: Phys. Rev. Lett. **101** (2008) 107006.
6) F. C. Hsu, J. Y. Luo, K. W. Yeh, T. K. Chen, T. W. Huang, P. M. Wu, Y. C. Lee, Y. L. Huang, Y. Y. Chu, D. C. Yan, and M. K. Wu: Proc. Natl. Acad. Sci. U.S.A. **105** (2008) 14262.
7) M. H. Fang, H. M. Pham, B. Qian, T. J. Liu, E. K. Vehstedt, Y. Liu, L. Spinu, and Z. Q. Mao: Phys. Rev. B **78** (2008) 224503.
8) K. W. Yeh, T. W. Huang, Y. L. Huang, T. K. Chen, F. C. Hsu, P. M. Wu, Y. C. Lee, Y. Y. Chu, C. L. Chen, J. Y. Luo, D. C. Yan, and M. K. Wu: Europhys. Lett. **84** (2008) 37002.
9) Y. Mizuguchi, F. Tomioka, S. Tsuda, T. Yamaguchi, and Y. Takano: arXiv:0811.1123.
10) W. Bao, Y. Qiu, Q. Huang, M. A. Green, P. Zajdel, M. R. Fitzsimmons, M. Zhernenkov, M. Fang, B. Qian, E. K. Vehstedt, J. Yang, H. M. Pham, L. Spinu, and Z. Q. Mao: Phys. Rev. Lett. **102** (2009) 247001.
11) S. Li, C. de la Cruz, Q. Huang, Y. Chen, J. W. Lynn, J. Hu, Y. L. Huang, F. C. Hsu, K. W. Yeh, M. K. Wu, and P. Dai: Phys. Rev. B **79** (2009) 054503.
12) C. de la Cruz, Q. Huang, J. W. Lynn, J. Li, W. Ratcliff II, J. L. Zarestky, H. A. Mook, G. F. Chen, J. L. Luo, N. L. Wang, and P. Dai: Nature **453** (2008) 899.
13) T. Nomura, S. W. Kim, Y. Kamihara, M. Hirano, P. V. Sushko, K. Kato, M. Takata, A. L. Shluger, and H. Hosono: Supercond. Sci. Technol. **21** (2008) 125028.
14) Q. Huang, Y. Qiu, W. Bao, M. A. Green, J. W. Lynn, Y. C. Gasparovic, T. Wu, G. Wu, and X. H. Chen: Phys. Rev. Lett. **101** (2008) 257003.
15) N. Ni, S. Nandi, A. Kreyssig, A. I. Goldman, E. D. Mun, S. L. Bud'ko, and P. C. Canfield: Phys. Rev. B **78** (2008) 014523.
16) J. Q. Yan, A. Kreyssig, S. Nandi, N. Ni, S. L. Bud'ko, A. Kracher, R. J. McQueeney, R. W. McCallum, T. A. Lograsso, A. I. Goldman, and P. C. Canfield: Phys. Rev. B **78** (2008) 024516.
17) Y. Mizuguchi, F. Tomioka, S. Tsuda, T. Yamaguchi, and Y. Takano: Appl. Phys. Lett. **94**

(2009) 012503.

18) B. C. Sales, A. S. Sefat, M. A. McGuire, R. Y. Jin, and D. Mandrus: Phys. Rev. B **79** (2009) 094521.

19) G. F. Chen, Z. G. Chen, J. Dong, W. Z. Hu, G. Li, X. D. Zhang, P. Zheng, J. L. Luo, and N. L. Wang: Phys. Rev. B **79** (2009) 140509.

20) M. H. Fang, B. Qian, H. M. Pham, J. H. Yang, T. J. Liu, E. K. Vehstedt, L. Spinu, and Z. Q. Mao: arXiv:0811.3021.

21) M. S. Torikachvili, S. L. Bud'ko, N. Ni, and P. C. Canfield: Phys. Rev. Lett. **101** (2008) 057006.

22) T. Park, E. Park, H. Lee, T. Klimczuk, E. D. Bauer, F. Ronning, and J. D. Thompson: J. Phys.: Condens. Matter **20** (2008) 322204.

23) P. L. Alireza, Y. T. Chris Ko, J. Gillett, C. M. Petrone, J. M. Cole, G. G. Lonzarich, and S. E. Sebastian: J. Phys.: Condens. Matter **21** (2009) 012208.

24) H. Kotegawa, H. Sugawara, and H. Tou: J. Phys. Soc. Jpn. **78** (2009) 013709.

25) H. Okada, K. Igawa, H. Takahashi, Y. Kamihara, M. Hirano, H. Hosono, K. Matsubayashi, and Y. Uwatoko: J. Phys. Soc. Jpn. **77** (2008) 113712.

26) Y. Mizuguchi, F. Tomioka, S. Tsuda, T. Yamaguchi, and Y. Takano: Appl. Phys. Lett. **93** (2008) 152505.

27) S. Margadonna, Y. Takabayashi, Y. Ohishi, Y. Mizuguchi, Y. Takano, T. Kagayama, T. Nakagawa, M. Takata, and K. Prassides: arXiv:0903.2204.

28) S. Medvedev, T. M. McQueen, I. Trojan, T. Palasyuk, M. I. Eremets, R. J. Cava, S. Naghavi, F. Casper, V. Ksenofontov, G. Wortmann, and C. Felser: arXiv:0903.2143.

29) D. Braithwaite, B. Salce, G. Lapertot, F. Bourdarot, C. Marin, D. Aoki, and M. Hanfland: J. Phys.: Condens. Matter **21** (2009) 232202.

30) H. Takahashi, K. Igawa, K. Arii, Y. Kamihara, M. Hirano, and H. Hosono: Nature **453** (2008) 376.

31) Y. Mizuguchi, F. Tomioka, S. Tsuda, T. Yamaguchi, and Y. Takano: arXiv:0810.5191.

32) A. Subedi, L. Zhang, D. J. Singh, and M. H. Du: Phys. Rev. B **78** (2008) 134514.

33) K. Murata, K. Yokogawa, H. Yoshino, S. Klotz, P. Munsch, A. Irizawa, M. Nishiyama, K. Iizuka, T. Nanba, T. Okada, Y. Shiraga, and S. Aoyama: Rev. Sci. Instrum. **79** (2008) 085101.

34) A. Kreyssig, M. A. Green, Y. Lee, G. D. Samolyuk, P. Zajdel, J. W. Lynn, S. L. Bud'ko, M. S. Torikachvili, N. Ni, S. Nandi, J. B. Leão, S. J. Poulton, D. N. Argyriou, B. N. Harmon, R. J. McQueeney, P. C. Canfield, and A. I. Goldman: Phys. Rev. B **78** (2008) 184517.

35) A. I. Goldman, A. Kreyssig, K. Prokeš, D. K. Pratt, D. N. Argyriou, J. W. Lynn, S. Nandi, S. A. J. Kimber, Y. Chen, Y. B. Lee, G. Samolyuk, J. B. Leão, S. J. Poulton, S. L. Bud'ko, N. Ni, P. C. Canfield, B. N. Harmon, and R. J. McQueeney: Phys. Rev. B **79** (2009) 024513.
36) T. Yildirim: Phys. Rev. Lett. **102** (2009) 037003.
37) J. N. Millican, D. Phelan, E. L. Thomas, J. B. Leão, and E. Carpenter: Solid State Commun. **149** (2009) 707.

**Figure captions**

Fig. 1. (Color online) Temperature dependence of the magnetization at magnetic field $B = 1$ T under various pressures of up to 1.40 GPa. The data include the magnetization arising from the pressure cell.

Fig. 2. (Color online) Temperature dependences of the electrical resistivity $\rho(T)$ (a) and the derivative with respect to temperature $d\rho/dT$ (b) under various pressures of up to 2.5 GPa, using the piston-cylinder-type cell. The $\rho(T)$ data at 0, 0.4, 1.1, 1.3, and 1.6 GPa are shifted upward by 3.5, 3.2, 2.5, 1.5, and 0.5 μΩm, respectively. The arrows indicate the characteristic temperatures $T_s$, $T_0$, and $T^*$ determined by the temperature at which $d\rho/dT$ is maximum.

Fig. 3. (Color online) Electrical resistivities at 4.2 and 200 K as a function of pressure.

Fig. 4. (Color online) Pressure dependences of the characteristic temperatures $T_s$, $T_0$, and $T^*$. The solid and open symbols indicate the data obtained by the electrical resistivity and magnetization measurements, respectively. The solid lines are guides to the eye.

Fig. 5. (Color online) Temperature dependence of resistance under various pressures of up to 19 GPa obtained using the DAC.

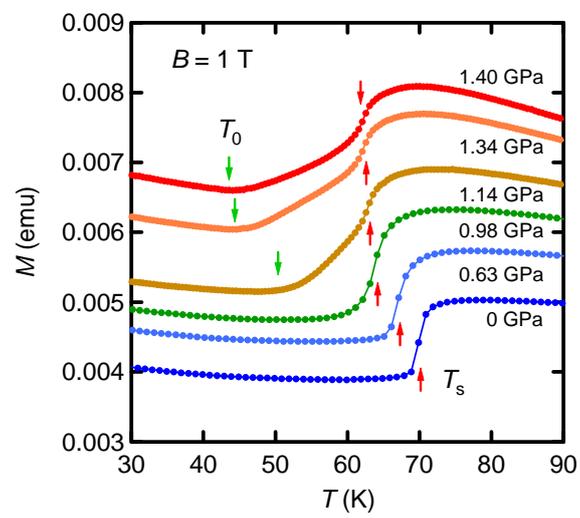

Fig. 1

H. Okada *et al.*

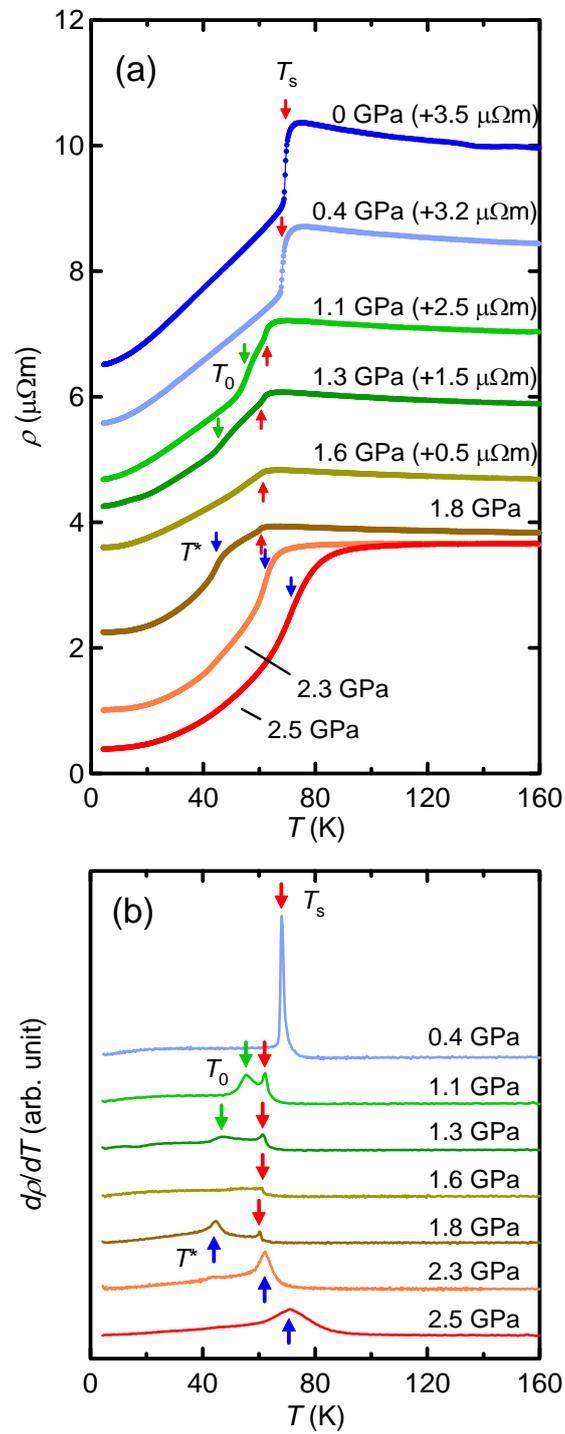

Fig. 2

H. Okada *et al*.

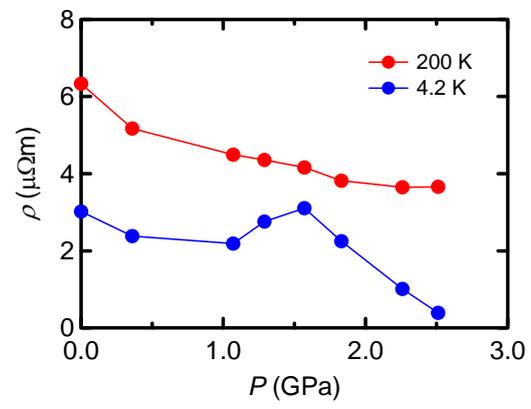

Fig. 3

H. Okada *et al.*

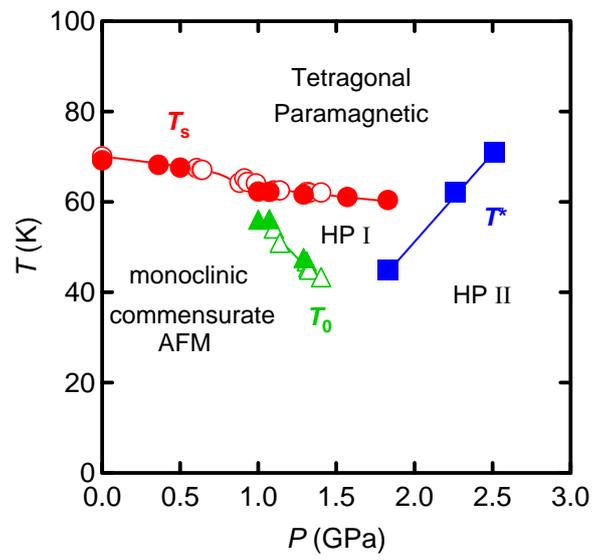

Fig. 4

H. Okada *et al.*

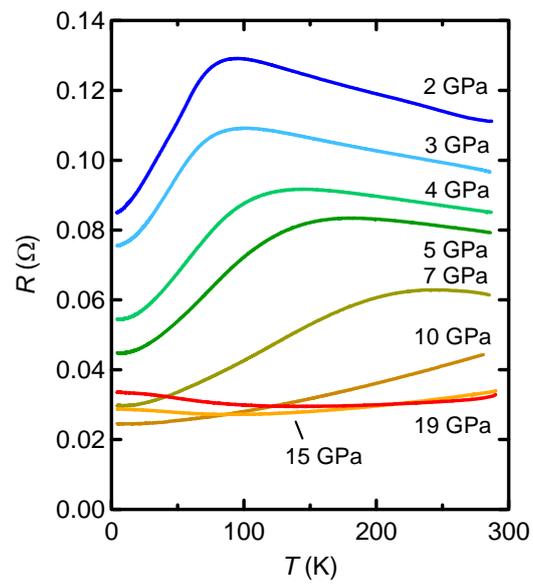

Fig. 5

H. Okada *et al.*